\documentclass[twocolumn,prl,showpacs]{revtex4}
\usepackage[dvips]{graphicx}
\usepackage{amsmath,amsfonts,amssymb,bm,ulem}
\begin{document}

\title{Strong localization of positive charge in DNA}
\author{D. B. Uskov and A. L. Burin}
\affiliation{Department of Chemistry, Tulane University, New Orleans LA, 70118}
\date{\today}
\begin{abstract}
Microscopic mechanisms of positive charge transfer in DNA remain unclear. A quantum state of electron hole in DNA is determined by the competition of a pi-stacking interaction $b$ smearing a charge between different base pairs and interaction $\lambda$ with the local environment which attempts to  trap charge. To determine which interaction dominates we investigated charge quantum states in various $(GC)_{n}$ sequences choosing DNA parameters satisfying  experimental data for the balance of charge transfer rates $G^{+} \leftrightarrow G_{n}^{+}$, $n=2,3$ \cite{FredMain}. We show that experimental data can be consistent with theory  only under an assumption $b\ll \lambda$ meaning that charge is typically localized within a single $G$ site. Consequently any DNA sequence including the one consisting of identical base pairs behaves more like an insulating material than a molecular conductor. 
\end{abstract}


\maketitle

{\bf 1.} 
Positive charge transfer in DNA is being extensively investigated since its experimental discovery \cite{FirstExperiment}, while to our knowledge there is no experimental demonstration of the negative charge transfer. Electron transfer in DNA can be responsible for the oxidative DNA damage \cite{FirstExperiment,Barton2,Barton3,GieseMB,Schuster1} and is possibly important for DNA repairing\cite{Raiskii,Taiwan}. Also an ability of DNA to promote long distant charge transfer can be used in molecular electronics \cite{MolecularWire1}. Therefore this problem is extensively investigated during the past decade but the microscopic mechanism of charge transfer is not understood yet.  We hope that this work may shed some light on this challenging problem. 

DNA contains two different sorts of base pairs $AT$ and $GC$ forming quasi-random sequences. The lowest ionization potential is attributed to a $GC$ pair (essentially G-base \cite{Sugiyama}).  Since the electron transfer  integral $b$ between adjacent bases does not exceed $AT$ - $GC$ ionization potential difference $b \leq \Delta \sim 0.5$eV, the quantum state of charge in a frozen environment will be localized near some $G$ base and its localization length is comparable to the interbase distance\cite{Shapiro}. Interaction with the environment breaks down this localization, inducing charge hopping between adjacent quantum states localized at $G$ sites. Indeed, according to experimental studies \cite{GieseMB,Schuster1} and theoretical model \cite{BBR1,BixonJortner} the sequence dependent charge transfer in DNA can be represented as the series of charge hops between adjacent $G$ bases serving as centers of localized states. An addition of $AT$ pair between adjacent $GC$ pairs dramatically reduces the charge hopping rate 
\cite{GieseMB,BixonJortner,BBR1} and therefore optimum base sequence for the most efficient charge transfer consists of identical base pairs. Therefore we are going to study a charge (hole) quantum state in sequences of identical $GC$ base pairs. 

The thermal energy at room temperature is very small compared to other characteristic energies of the problem so we  can use the ground state of  the hole coupled to environment as a representative state. Then the spatial size of the hole state is determined by the competition of the delocalization of charge due to the pi-stacking interaction of heterocyclic groups belonging to adjacent bases and the localization caused by the environment polarization around the charge. The charge delocalization energy can be characterized by the effective electron transfer integral $b$ and the localization energy is given by the medium reorganization energy $\lambda$\cite{Igor,abReview,LeBard,Voityuk,Berashevich}. Delocalization of charge over $k$ base pairs leads to the gain in the energy $E_{del}\sim -2b+b/k^2$, while the reorganization energy scales with the size of charge wavefunction as $E_{loc}\sim -\lambda/k$ (see e. g. \cite{Igor}). Optimization of the  total energy with respect to the number of sites $k$ estimates the size of the ground state charge wavefunction $k\sim 2b/\lambda$. At zero temperature the hole is localized, while at finite temperature it can hop to different states because of its interaction with the fluctuating environment. In the translationally invariant system ($(GC)_{n}$ or $(AT)_{n}$) the potential barrier separating two configurations can be estimated as the energy price for increasing the size of the wavefunction by one more site $k\rightarrow k+1$ compared to its optimum state. If $b\ll \lambda$ then this energy is given by the half of the reorganization energy 
$\Delta \approx \lambda/2 \gg k_{B}T$, 
while in the opposite limit we get a very small value $\Delta\sim \lambda^{4}/(16b^{3})$ which quickly becomes negligible at moderately large $b/\lambda$. 

It is important to determine the true relationship of $b$ and $\lambda$ in DNA. Indeed, in the regime $b\gg \lambda$ a DNA molecule made of identical base pairs would behave as a one-dimensional conductor, while in the opposite limit DNA always acts like an insulator. Existing estimates in literature \cite{Igor,abReview,LeBard,Voityuk,Conwell,Berashevich,Sugiyama} do not help much in resolving this issue, because there is a large controversy between different approximations. In particular, various estimates for the electron transfer integral $b$ range from $0.05$eV\cite{Voityuk} to $0.5$eV \cite{Sugiyama}, and all calculations of $b$ ignore vibrational rearrangements\cite{comment1}. The estimates for the reorganization energy $\lambda$ using the continuous medium approach range from $0.25$eV\cite{Berashevich} to more than $1$eV \cite{Igor} due to uncertainty in the water dielectric constant value near the DNA molecule. Therefore based on different approaches different conclusions are made about the charge state including a prediction of a propagating intermediate size polaron  \cite{Conwell,Berashevich,Sugiyama} or a small radius polaron essentially localized within a single base\cite{Voityuk}. 

To resolve this problem we suggest an alternative method to study the charge quantum state within the DNA molecule, using experimental data sensitive to the relationship of two key parameters of the theory, $b$ and $\lambda$. Namely, we exploit the rate constants for the balance of charge transfer rates between different $(GC)_{n}$ complexes, measured by Lewis and coworkers \cite{FredMain}
\begin{equation} \label{GrindEQ__5_} \begin{array}{l} {G^{+} +GG\mathop{\rightleftarrows }\limits_{k_{-t}^{GG} }^{k_{t}^{GG} } G+(GG)^{+}, \, \, \, \, \frac{k_{t}^{GG}}{k_{-t}^{GG}}=7.7\pm 1} \\ {G^{+} +GGG\mathop{\rightleftarrows }\limits_{k_{t}^{GGG} }^{k_{t}^{GGG} } G+(GGG)^{+}, \, \, \, \,\frac{k_{t}^{GGG}}{k_{-t}^{GGG}}=20\pm 1}. \end{array} \end{equation}
In the thermal equilibrium these ratios are determined by the base pair  partition functions 
\begin{eqnarray}
r_{2}=\frac{k_{t}^{GG}}{k_{-t}^{GG}}=\frac{Z_{2+}Z_{1}}{Z_{2}Z_{1+}}, 
\nonumber\\ 
r_{3}=\frac{k_{t}^{GGG}}{k_{-t}^{GGG}}==\frac{Z_{3+}Z_{1}}{Z_{3}Z_{1+}}. 
\label{eq:rat1}
\end{eqnarray}
where $Z_{n+}$ stands for the partition function for $G_{n}$ sequence containing a single hole in it, while $Z_{n}$ is the partition function for the same base sequence, but without the hole.  Since $G$ and $G_{n}$ complexes are separated by some $AT$ bridge we can ignore their interactions.   

\begin{figure}[b]
\centering
\includegraphics[width=9cm]{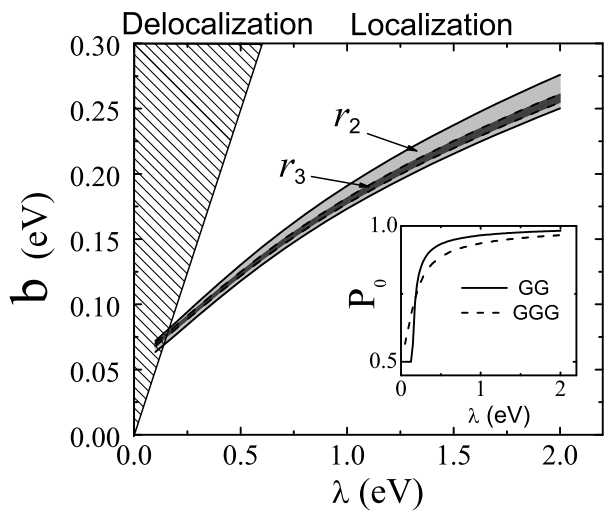}
\caption{The domains consistent with the experimental ratios of reaction rates Eq. (\ref{GrindEQ__5_}), dark grey for $GGG$ and  light grey for $GG$. Inset shows the fraction of the quantum charge state belonging to the central site ($P_{0}$) vs. the reorganization energy $\lambda$. 
\label{fig:potential} }
\end{figure}

Both ratios in Eq. (\ref{GrindEQ__5_}) depend on two parameters $b$ and $\lambda$ and the thermal energy at room temperature 
$k_{B}T \sim 0.026eV$. Below we calculate both ratios using tight binding model for $G_{n}$ complexes and standard linear response theory for  charge interaction with the environment \cite{abReview,Marcus}. Theory determines the domain of parameters $\lambda$ and $b$ satisfying experimental data Eq. (\ref{GrindEQ__5_}). {\it We demonstrate that any choice of $\lambda$ and $b$, satisfying Eq. (\ref{GrindEQ__5_}), corresponds to the regime $b \ll \lambda$, where the hole in its ground state is localized essentially in a single $G$ base (see Fig. \ref{fig:potential}). }

{\bf 2.} 
The chain of $n$ $GC$ base pairs can be described by the tight binding Hamiltonian coupled to the classical environment represented by coordinates $X_{i}$, $i=1, ...n$ coordinates for each DNA site 
\begin{eqnarray}
\widehat{H} = \widehat{H}_{hole} + \widehat{H}_{media} + \widehat{V}_{int},  
\nonumber\\
\widehat{H}_{hole}= -b\sum_{i=1}^{n-1}(c_{i}^{+}c_{i+1}+c_{i+1}^{+}c_{i}), 
\nonumber\\ 
\widehat{H}_{media} = \frac{1}{2\lambda}\sum_{i=1}^{n}X_{i}^{2}, \, \, \,
\widehat{V}_{int} = -\sum_{i=1}^{n}X_{i}c_{i}^{+}c_{i}. 
\label{eq:Hamiltonian}
\end{eqnarray}
Here $c_{i}$, $c_{i}^{+}$ are operators of creation and annihilation of electron hole in a site $i$. Classical coordinates $X_{i}$ describing the polar environment are directly coupled to the local charge density $n_{i}=c_{i}^{+}c_{i}$. The solvent energy is  expressed as a bilinear form with respect to solvent coordinates, which is justified by a standard assumption that polarization fields are small compared to atomic fields\cite{Marcus} so we can ignore $X^{3}$ terms. We assume that only classical degrees of freedom with excitation energy comparable or less than the thermal energy are left in Eq. (\ref{eq:Hamiltonian}), while high energy modes are integrated out. This may lead to the renormalization of parameters in the system Hamiltonian Eq. (\ref{eq:Hamiltonian}) (see e. g. \cite{comment1,ab_prl}) and we assume that this renormalization is made. We do not include off-diagonal terms $X_{i}X_{j}$, $i\neq j$ into the Hamiltonian. This is justified because they are smaller than the diagonal ones \cite{LeBard}. It can be shown that for $G_{2}$ sequence the problem including off-diagonal terms can be reduced to the diagonal model Eq. (\ref{eq:Hamiltonian}) with the replacement of the single site reorganization energy $\lambda$ with the reorganization energy for charge transfer between adjacent sites. For $GGG$ sequence the similar replacement with removal off-diagonal terms remains a good approximation\cite{promise}. Note that the addition of $A$ or $T$ bases surrounding $G_{n}$ sequences leads to small changes in our results because the electron transfer integral is smaller in average between $A$ and $G$ base then between two $G$ bases and also because of the large ionization potential difference of $A$ and $G$ bases \cite{Voityuk}. Also we ignore the second $C_{n}$ strand because of the weak coupling between strands \cite{Voityuk} and large difference of their ionization potentials\cite{Sugiyama}. 

We study the ratios of charge transfer rates Eq. (\ref{eq:rat1}). Each partition function is given by $Z_{n}=\int dX_{1}...dX_{n}Tr e^{-\beta H_{n}}$, where $H_{n}$ is $G_{n}$ Hamiltonian Eq. (\ref{eq:Hamiltonian}), trace is taken only over the states with the single hole ($Z_{n+}$) or no holes ($Z_{n}$) and $\beta=1/(k_{B}T)$.  If there is no hole  calculations are reduced to multiple evaluation of a Gaussian integral leading to 
\begin{eqnarray}
Z_{n} = c^{n}, \, \, \, c=\sqrt{\frac{2\pi\lambda}{\beta}}.  
\label{eq:zero_part_trace}
\end{eqnarray}     
For the sequences containing a hole an analytical expression can be obtained only for $n=1$ 
\begin{eqnarray}
Z_{1+} = ce^{\beta\lambda/2}.  
\label{eq:one_part_trace1}
\end{eqnarray} 
For $n=2$, $3$ one can perform analytical integration over a ``center of mass'' coordinate $X_{1}+..X_{n}$ which is coupled to the conserving operator of the total number of particles $c_{1}^{+}c_{1}+..+c_{n}^{+}c_{n}=1$. This reduces the calculations to single and double integrals, respectively. Below we give the expression for $(GG)^{+}$ partition function   
\begin{eqnarray}
 Z_{2+} = (\sqrt{2}c)e^{\beta\lambda/4}\int_{-\infty}^{+\infty}du e^{-\frac{\beta u^{2}}{4\lambda}}\cosh\left(\beta\sqrt{\frac{u^{2}}{4}+b^{2}}\right), 
\label{eq:one_part_trace2}
\end{eqnarray} 
while the expression for $Z_{3+}$ is more complicated and will be published elsewhere\cite{promise}.

{\bf 3}. 
We have performed numerical evaluations of ratios in Eq. (\ref{eq:rat1}) to find domains of parameters $b$ and $\lambda$ satisfying Eq. (\ref{GrindEQ__5_}) and show these domains in Fig. \ref{fig:potential}. The upper (lower) border of each domain is defined by the maximum (minimum) value of ratios $r_{2}$ and $r_{3}$ Eq. (\ref{eq:rat1}) within the experimental error ($8.7$ ($6.7$) for $GG$ and $21$ ($19$) for $GGG$). The acceptable domain of parameters for the $GGG$ sequence fully belongs to the corresponding domain for the $GG$ sequence. Thus the domains for $GG$ and $GGG$ base sequences are completely consistent with each other. Therefore we cannot determine parameters $\lambda$ and $b$ better then using the ``dark'' domain for $GGG$. This information is still sufficient to consider the localization of the hole wavefunction in $G_{n}$ aggregates.
  
Since the thermal energy $k_{B}T \approx 0.026$eV is smaller than other characteristic energies of the system (remember that the minimum estimate for the reorganization energy is $\lambda \sim 0.25$eV \cite{Berashevich}) we can characterize the wavefunction using the system ground state at coordinates $X_{i}$ minimizing the ground state energy. In the relevant domain of parameters in Fig. \ref{fig:potential} ($\lambda > 0.25$) the ground state wavefunction is centered at one of $G$ bases (left or right ones for the $GG$ sequence and the central one for a $GGG$ sequence). We characterize this state by the probability $P_{0}$ for the particle in the ground state to be in this central site. One can show that $P_{0}=X_{i}/\lambda$, where $i$ is the central site\cite{promise}.  

For instance, for the $GG$ sequence the expression for the hole ground state energy at arbitrary coordinates $X_{1}$, $X_{2}$ reads  
$E_{2}=\frac{X_{1}^{2}+X_{2}^{2}}{2\lambda}-\frac{X_{1}+X_{2}}{2}-\sqrt{\frac{(X_{1}-X_{2})^2}{4}+b^2}$.
In the regime of interest $2b<\lambda$ (see Fig. \ref{fig:potential}) the minimum of energy is given by 
\begin{equation}
E_{2min}=-\lambda/2-b^2/\lambda,   
\label{eq:pairenergymin}
\end{equation}
and it is realized at $X_{1}=\lambda/2\pm\sqrt{(\lambda/2)^2-b^2} = \lambda-X_{2}$. Accordingly 
$P_{0}=\frac{X_{1}}{\lambda}=\frac{\lambda+\sqrt{\lambda^2-4b^2}}{2\lambda}.$ 
Note that if $2\lambda<b$ the ground state wavefunction is symmetric in the minimum $X_{1}=X_{2}=\lambda/2$ and the energy of this state is given by  
\begin{equation}
E_{2symm}=-\lambda/4-b.  
\label{eq:pairenergyminsaddle}
\end{equation}
In the case of $2b<\lambda$ this symmetric state is the transition state (saddle point in the energy function $E_{2}(X_{1},X_{2})$) between the energy minima centered in the first and the second $G$'s.

For the $GGG$ complex the probability to reside in the central site was evaluated numerically. Both probabilities $P_{0}$ obtained for the $\lambda$ - $b$ line corresponding to the ratio $r_{2}=7.7$ Eq. (\ref{eq:rat1}) are shown in the inset in Fig. \ref{fig:potential}. It is clear from this graph that for both $GG$ and $GGG$ sequences the wavefunction of the hole is essentially localized in the single $G$ site. For instance at the minimum value of $\lambda \sim 0.25eV$ we have $85\%$ and $78\%$ of the probability to find the particle in that site for $GG$ and $GGG$ sequences, respectively.  These probabilities increase with increasing of the reorganization energy to $1$eV up to $96\%$ and $94\%$, respectively. 

Thus we come to the important conclusion that wavefunctions of hole are essentially localized in the single $G$ site for $G_{n}$ sequences. This conclusion differs from the predictions of previous work \cite{Berashevich,Conwell,Taiwan} where the polaron of an intermediate range was used to describe the quantum state of the hole. Our conclusion is justified by the agreement with the experiment\cite{FredMain}.

{\bf 4}. An impressive consistency between $GG$ and $GGG$ in Fig. \ref{fig:potential} is not accidental and can be explained by the strong localization of charge wavefunctions. In the regime of strong localization the partition function $Z_{n+}$ for $n\geq 2$ consists of $n$ contributions of energy minima corresponding to wavefunctions centered in all $n$ $G$ sites with coordinates $X$ realizing the corresponding energy 
minimum $X_{i} \approx \lambda \gg X_{k}$, $k\neq i$ for the state centered at site $i$. Since in the zero order approximation in $b/\lambda$ each quantum state is localized at one site one can neglect the difference in preexponential factors for the case of $b=0$ and we can approximate the partial $i^{th}$ contribution to the partition function as $Z_{n}^{i}=c^{n}e^{-\beta E_{i}}$ (see Eq. (\ref{eq:zero_part_trace})), where $E_{i}$ is the energy of the ground state for coordinates $X$ realizing the local minimum. First order correction to the energy $E_{i}^{(0)} = -\lambda/2$ is important because it is in the exponent and multiplied by the large factor $\beta$. For the two states at the edges this correction can be expressed as $E_{1} = E_{n} \approx -\lambda/2-b^{2}/\lambda$, which coincides with the ground state energy for $GG$ Eq. (\ref{eq:pairenergymin}). This is not surprising because the contributions of non-neighboring sites is negligible due to the strong localization of charge. For $n-2$ remaining states the correction to the energy should be doubled because of the addition of contributions of two neighbors so we got $E_{i}^{(1)} = -\lambda/2 - 2b^{2}/\lambda$, $1<i<n$. 
Consequently, we can approximate the ratio $r_{n}$ (cf. Eq. (\ref{eq:rat1})) as 
\begin{eqnarray}
r_{n}=Z_{n+}Z_{1}/(Z_{n}Z_{1+}) \approx 2e^{\beta b^2/\lambda}+(n-2)e^{2\beta b^2/\lambda}. 
\label{eq:theorGGG} 
\end{eqnarray}
Particularly, one can show that  $(Z_{3+}Z_{1}/(Z_{3}Z_{1+}) \approx (Z_{2+}Z_{1}/(Z_{2}Z_{1+})+((Z_{2+}Z_{1}/(2Z_{2}Z_{1+}))^2)$ and this relationship is satisfied for the experimental values of ratios within the accuracy of the experiment. This explains the consistency of domains for $GG$ and $GGG$ Fig. \ref{fig:potential}. Using Eq. (\ref{eq:theorGGG}) one can predict that ratios $r_{n}$ form arithmetic series. Particularly for the balance between $G$ and $G_{4}$ sequence we predict the ratio $r_{4}=2r_{3}-r_{2}=32.3$. This estimate agrees with our numerical calculations for the $G_{4}^{+}$ partition function.  One should notice that in case of delocalization $b>\lambda/2$ the difference in energy between $G^{+}$ and $G_{2}^{+}$ is $b\geq 0.15eV$ and one would expect $r_{2} \geq e^{\frac{b}{k_{B}T}} \geq 300$ which is not the case in the experiment.

{\bf 5.} 
Thus we considered the quantum state of the positive charge (hole) in poly-$G$ - poly-$C$ base sequence. It turns out that the agreement with the experimental data for the ratios $r_{n}$ Eq. (\ref{eq:rat1}) for $n=2$, $3$ can be attained only assuming the strong localization of charge within almost a single $G$-base.  The charge in DNA then behaves as a small polaron with the size less than the interbase distance. Based on our theory we predict all other ratios $r_{n}=7.7+12.3\cdot (n-2)$.   Notwithstanding this prediction we are not able to identify more accurately the  electron transfer integral $b$ and the reorganization energy $\lambda$ using experimental data only for the ratios. 

Based on our theory we can suggest to find these parameters measuring the temperature dependence of the charge transfer rate through poly-$G$ - poly-$C$ base sequence. We expect that this temperature dependence will be described by the Arrhenius law with the activation energy defined by the difference of charge symmetric transition state energy within $(GG)^{+}$ base pair Eq. (\ref{eq:pairenergyminsaddle}) and the charge ground state energy for $(GG)^{+}$ state Eq. (\ref{eq:pairenergymin}) $E_{A}=\lambda/4-b+b^{2}/\lambda$. 

This work is supported by the NSF CRC Program, Grant No. 0628092.
The authors acknowledge Fred Lewis, Michael Wasielewski, George Schatz, Thorsten Fiebig and Mark Ratner for useful discussions.


\end{document}